\documentclass[12pt,notitlepage,a4paper]{article}
\pdfoutput=1
\usepackage{color,graphicx}
\usepackage{amssymb}
\usepackage{delarray,amsmath,bbm}
\usepackage[latin1]{inputenc}
\usepackage[american]{babel}
\usepackage{cite}
\usepackage{subfigure}
\usepackage{epsfig}

\newcommand{\be}{\begin{equation}}
\newcommand{\ee}{\end{equation}}
\newcommand{\beqa}{\begin{eqnarray}}
\newcommand{\eeqa}{\end{eqnarray}}
\numberwithin{equation}{section}

\newfont{\namefont}{cmr10}
\newfont{\addfont}{cmti7 scaled 1440}
\newfont{\boldmathfont}{cmbx10}
\newfont{\headfontb}{cmbx10 scaled 1728}

\begin{document}
\baselineskip=15.5pt
\pagestyle{plain}
\setcounter{page}{1}

\begin{center}
\rightline{IFT-UAM/CSIC-16-039 }
\vspace{2cm}

\renewcommand{\thefootnote}{\fnsymbol{footnote}}

\begin{center}
{\Large \bf  Holographic Quenches with a Gap}
\end{center}
\vskip 0.1truein
\begin{center}
\vspace{6mm}
\bf{Emilia da Silva${}^1$, 
Esperanza Lopez${}^1$, 
Javier Mas${}^2$ and Alexandre Serantes${}^2$ }
\end{center}
\vspace{0.5mm}

\begin{center}\it{
${}^1$ Instituto de F\'\i sica Te\'orica IFT UAM/CSIC \\
Universidad Aut\'onoma de Madrid\\
 28049 Cantoblanco, 
 Madrid, Spain
}
\end{center}

\begin{center}\it{
${}^2$Departamento de  F\'\i sica de Part\'\i  culas \\
Universidade de Santiago de Compostela, 
and \\
Instituto Galego de F\'\i sica de Altas Enerx\'\i as IGFAE\\
E-15782 Santiago de Compostela, Spain}
\end{center}

\vspace{0.5mm}

\setcounter{footnote}{0}
\renewcommand{\thefootnote}{\arabic{footnote}}

\let\thefootnote\relax\footnotetext{Emails: emilia.dasilva@csic.es, esperanza.lopez@uam.es, javier.mas@usc.es, \\ alexandre.serantes@gmail.com}

\vspace{0.6in}

\begin{abstract}
\noindent 
In order to holographically model quenches with a gapped final hamiltonian, we consider a gravity-scalar theory in anti-de Sitter space with an infrared hard wall. We allow a time dependent profile for the scalar field at the wall. This induces an energy exchange between bulk and wall and generates an oscillating scalar pulse. We argue that such backgrounds are the counterpart of quantum revivals in the dual field theory. We perform a qualitative comparison with the quench dynamics of the massive Schwinger model, which has been recently analyzed using tensor network techniques. Agreement is found provided the width of the oscillating scalar pulse is inversely linked to the energy density communicated by the quench. We propose this to be a general feature of holographic quenches.

\end{abstract}

\smallskip
\end{center}
\newpage


\section{Introduction}

Holography has proven useful for the study quantum field theories in regimes where standard techniques fail. Besides strong coupling physics, the out of equilibrium dynamics of quantum systems has emerged in the last years as a topic where holographic techniques can provide important insight.

The dictionary between classical geometries and field theory states brings into correspondence black holes in an asymptotically anti-de Sitter space with states of thermal equilibrium. It is then natural to interpret processes of gravitational collapse in terms of evolution towards thermal equilibrium in the dual field theory \cite{Banks:1998dd,Danielsson:1999fa}. Under certain conditions matter distributions give rise to oscillating geometries, which might end in the formation of a horizon on longer time scales or keep bouncing forever \cite{Bizon:2011gg,Buchel:2012uh}. Following the previous logic, this opens the door to describe holographically excited states in a strongly interacting theory which do not lead to fast thermalization \cite{Abajo-Arrastia:2014fma,daSilva:2014zva}. Although fast thermalization is generically expected, a variety of systems and initial conditions are known where it is not realized \cite{Polkovnikov:2010yn}. It is exciting that holography might help to address this question, which is a a subject of active research in condensed matter and quantum information theory.

We are interested in holographically modeling quantum quenches where the final hamiltonian is gapped. We will use for that a very simple geometric setup consisting in introducing an infrared hard wall in AdS. Gravitational collapse of a massless scalar field in the hard wall model has been considered before \cite{Craps:2013iaa,Craps:2014eba}. New ingredients are however introduced. We show that a mass can be assigned to the wall which affects the infrared physics of the dual field theory. The space of couplings is parameterized by this mass together with the value of the scalar field at the wall. A time dependent profile for the scalar field at the wall allows to move in the space of couplings while introducing a homogeneous excitation in the system. In this way a quench is holographically simulated, where initial and final hamiltonian differ.

Tensor network techniques have been applied very successfully to the study of ground states properties \cite{Orus:2013kga}. More recently they have been used to study real time dynamics, providing examples with which to crosscheck the holographic results. The response of the massive Schwinger model to a quench was addressed in \cite{Buyens:2013yza}. In the limit of small energy quenches the resulting evolution exhibits quantum revivals. Namely, the initial out of equilibrium states is periodically reconstructed after intermediate stages of apparent decoherence \cite{Robinett2004}. We compare the revival phenomenology of the massive Schwinger model with the oscillating geometries generated by a scalar pulse in the hard wall mode, finding consistency.

The plan of the paper is the following. In Section 2 we introduce the hard wall model. In Section 3 we study the infrared physics of its dual field theory. Time dependent boundary conditions at the wall are described in Section 4. In Section 5 we identify the criteria for the hard wall model to provide a good qualitative description of a quench, taking the Schwinger model as guideline. We end in Section 6 with a discussion on open issues.

\section{The hard wall model}

We want to model a 1+1 dimensional QFT on an infinite line, hence we will work in AdS$_3$ with Poincare slicing. A crude holographic representation of a mass gap is achieved by imposing an infrared cutoff in the bulk, modeled as an infinitely hard wall. Such simple set up will yet allow us to get interesting insight into the holographic dictionary. 

In the hard wall framework, we will consider gravity coupled to a massless scalar field \cite{Craps:2013iaa,Craps:2014eba}. Configurations with translational invariance along the boundary spatial direction can be described using the diagonal ansatz for the metric 
\be
ds^2 = \frac{1}{z^2}\left( - A(t,z) e^{-2\delta(t,z)} dt^2 + {dz^2 \over A(t,z)} +  d x^2\right) \, .
\label{metric}
\ee
The AdS boundary corresponds to $z\!=\!0$, while the location of the infrared wall will be denoted by $z_0$. The equations of motion are
\beqa
\dot\Phi &=& \left( A e^{-\delta} \Pi\right)' ~~~,~~~~~~\dot \Pi = z\left({ A e^{-\delta} \Phi \over z}\right)' \, ,\label{eqforphi} \\[1mm]
\delta' &=& {z \over 2} \,(\Phi^2 + \Pi^2) ~~~,~~~  A'={z \over 2} \, A\,  (\Phi^2 + \Pi^2)+ {2 \over z} (A-1) \label{eqforA} 
\eeqa
where $\Phi\!=\!\phi'$ and $\Pi\!=\!A^{-1} e^\delta \dot{\phi}$, with $\phi'$ and  $\dot{\phi}$ the space and time derivatives of the scalar field respectively. 

The equations of motion allow an inflow of energy from the AdS boundary into the bulk, triggered by the variation of the non-normalizable component of the scalar field. 
We set this component to zero along the paper by requiring the scalar field to vanish asymptotically. This insures the conservation of the total mass 
\be
M=z_0^{-2} (1-A_0) + {1 \over 2} \int_0^{z_0} y^{-1} (\Phi^2 + \Pi^2) A \, dy \; .
\label{mass}
\ee
The value of the metric function $A$ at the wall, $A_0$, is determined by the momentum constrain 
\be
\dot A = z \, A^2 e^{-\delta} \Phi \Pi \, .
\label{momconst}
\ee
We assume that the position of the infrared wall is a parameter of the model not susceptible of any variation. Under this condition, $A_0$ can only be constant if the scalar field satisfies Dirichlet or Neumann boundary conditions at the wall  \cite{Craps:2013iaa,Craps:2014eba}. In the following we introduce a freedom in the hard wall model which has not been addressed before. We will allow for a time dependent profile of the scalar field at the wall. In particular, our seed for the evolution will be a non-trivial function
\be
\Pi_0(t)=\Pi(t,z=z_0) \, .
\label{bc}
\ee
The momentum constraint ensures the conservation of the total mass also under these general boundary conditions. The first term on the {\it rhs} of \eqref{mass} can be considered a property of the wall, which we will refer to as $M_0$, the mass of the wall, while the second is the contribution of the scalar profile to the mass. A time dependent scalar profile at the wall induces an energy exchange between the gravity sector and the scalar field, represented respectively by the first and second terms in \eqref{mass}.  

A vanishing value of the wall mass corresponds to introducing an infrared cutoff in empty AdS space. Positive values $M_0 z_0^2\!<\!1$ relate to black hole geometries whose horizon lies behind the wall. Negative wall masses are linked to geometries which in the absence of the wall contain a naked singularity at the origin. The wall excises the singularity, leaving open their consistency as holographic backgrounds. 

A trivial case of \eqref{bc} are Dirichlet boundary conditions. The equations \eqref{eqforphi}-\eqref{eqforA} with Dirichlet boundary conditions at the wall admit static solutions with a non-trivial scalar profile \cite{Craps:2014eba}. The time independent Klein-Gordon equation simplifies to
\be
z^{-1} A e^{-\delta} \Phi = \alpha \; ,
\label{cmotion}
\ee
where $\alpha$ is an integration constant.
The metric \eqref{metric} allows the gauge freedom of reparameterizations along the time coordinate. Using this, $t$ can be fixed to be the proper time at the boundary, which is the natural choice for holographic applications. The integration constant $\alpha$ is then identified with the expectation value of the field theory operator dual to the scalar field, $\alpha=2 \langle {\cal O} \rangle$.

\begin{figure}[h]
\begin{center}
\includegraphics[width=6.5cm]{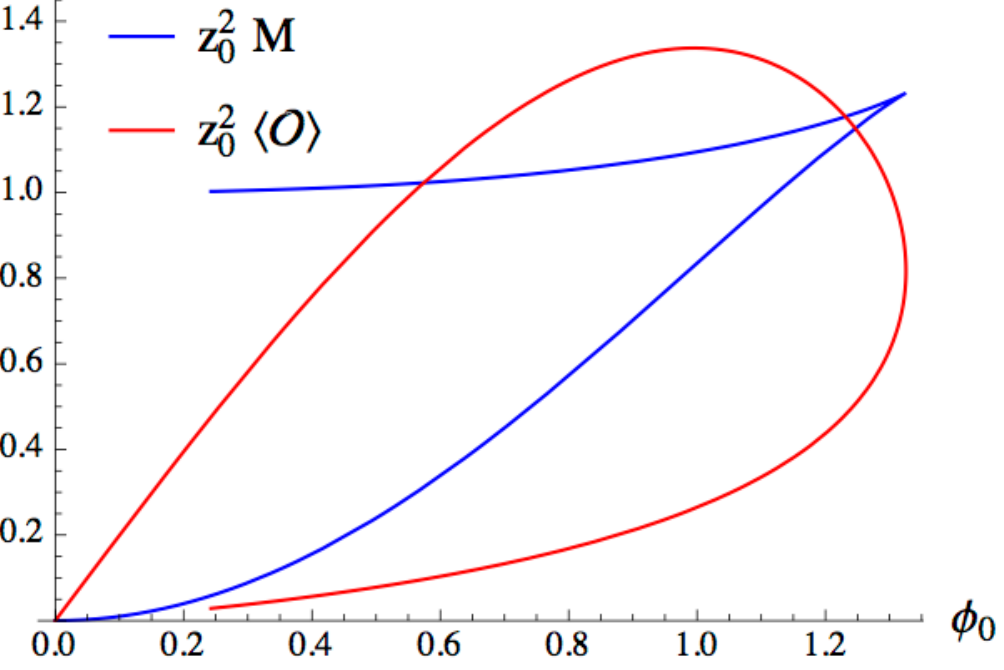}~~~
\includegraphics[width=7cm]{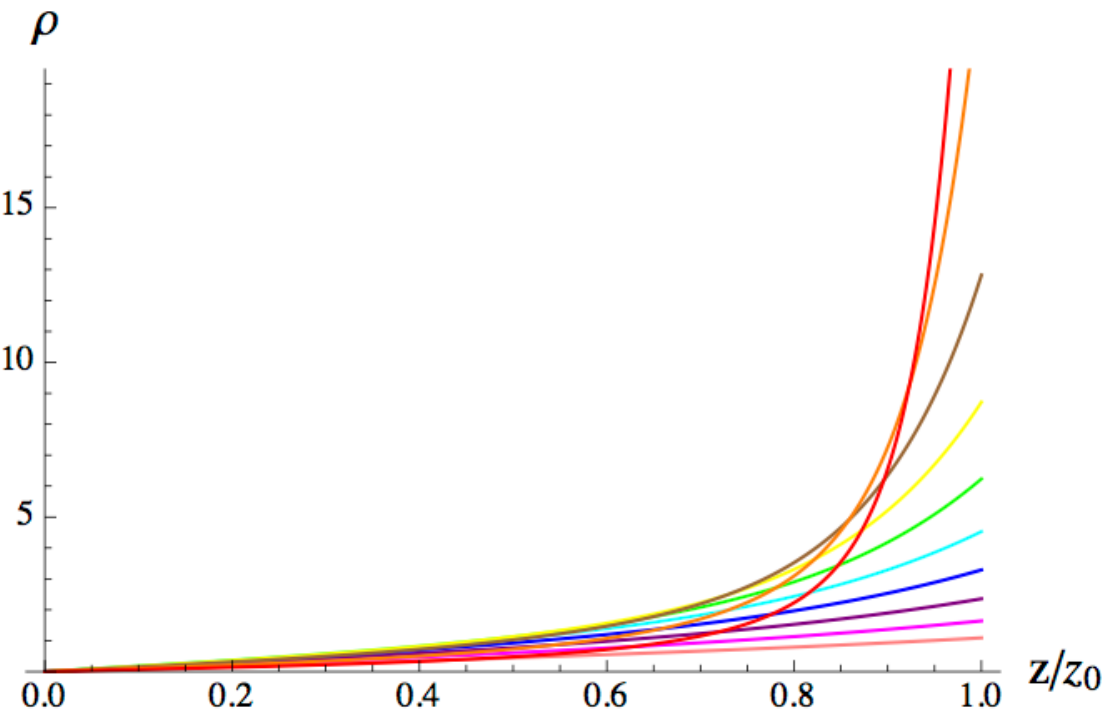}
\end{center}
\caption{\label{fig:static} Left: Properties of static scalar configurations with vanishing wall mass and Dirichlet boundary conditions $\phi(z_0)\!=\!\phi_0$. Right: Radial energy profile of solutions along the stable branch for $\phi_0\!=\!.5,.6,..,1.3$ and the threshold value $\phi_0\!=\!1.32$ from bottom to top.}
\end{figure}

As an example, the set of static solutions without horizons with vanishing wall mass is plotted in Fig.\ref{fig:static}a; other values of $M_0$ behave similarly. 
Static solutions exists up to a threshold for $\phi_0$, the scalar field at wall. There are two branches of solutions joining at it. Those with higher mass are linearly unstable. They approximate a black hole solution with horizon at the position of the wall as $\phi_0$ decreases. Consequently $M z_0^{2}$ tends to unity and $\langle {\cal O} \rangle$ to zero along the unstable branch. The lower mass branch is stable. Remarkably it includes regular solutions whose Schwarzchild radius is neatly above the wall position. We will use the integrand in \eqref{mass}
\be
\rho={1 \over 2 z}   (\Phi^2 + \Pi^2) A \; ,
\ee
to describe the radial energy distribution of the scalar configuration. Fig.\ref{fig:static}b shows this function for several values of $\phi_0$ along the stable branch.

\section{Infrared physics}

We analyze the dual field theory infrared physics associated to the static solutions by focussing on two quantities: {\it i)} the correlation length, defined as the size of the interval that saturates the large N contribution to the entanglement entropy; and {\it ii)} the mass gap, defined as the lowest normal frequency over a static solution without horizon. 

The entanglement entropy of a region is defined as the von Neumann entropy of the density matrix traced over the degrees of freedom in its complementary. The entanglement entropy of an interval is measured holographically by the length of a geodesic that anchors at the AdS boundary on the interval endpoints \cite{Ryu:2006bv}-$\!$\cite{Hubeny:2007xt}, and is homologous to it \cite{Fursaev:2006ih}. We will introduce a modification of the second condition in the hard wall model. Besides geodesics which satisfy the homology constraint, we also allow geodesics which together with their associated AdS boundary interval are homologous to an interval at the wall. In mathematical terms this means that the geodesic should be homologous to its associated interval in the relative homology defined with respect to the infrared boundary. The relaxation of the homology constraint is necessary because geodesics which intersect the AdS boundary at the endpoints of a sufficiently large interval unavoidably intersect the infrared wall. 

A question arises on the possible contribution of the length of the interval at the wall to the entanglement entropy. Including it would lead to an extensive behavior of the entanglement entropy, which is characteristic of a thermal state or a highly excited state which violates the area law. The eigenfrequencies of the laplacian in the dual geometries should have an imaginary component, reflecting the finite lifetime of excitations in such field theory states. This is verified in the black hole backgrounds, dual to thermal equilibrium. However, the frequencies of the harmonic modes in the hard wall setup with Dirichlet or Neumann boundary conditions at the wall are purely real. Hence in order to obtain a consistent infrared physics we must assume that the interval at the wall does not play a role in the computation of the holographic entanglement entropy. 

\begin{figure}[h]
\begin{center}
\includegraphics[width=12cm]{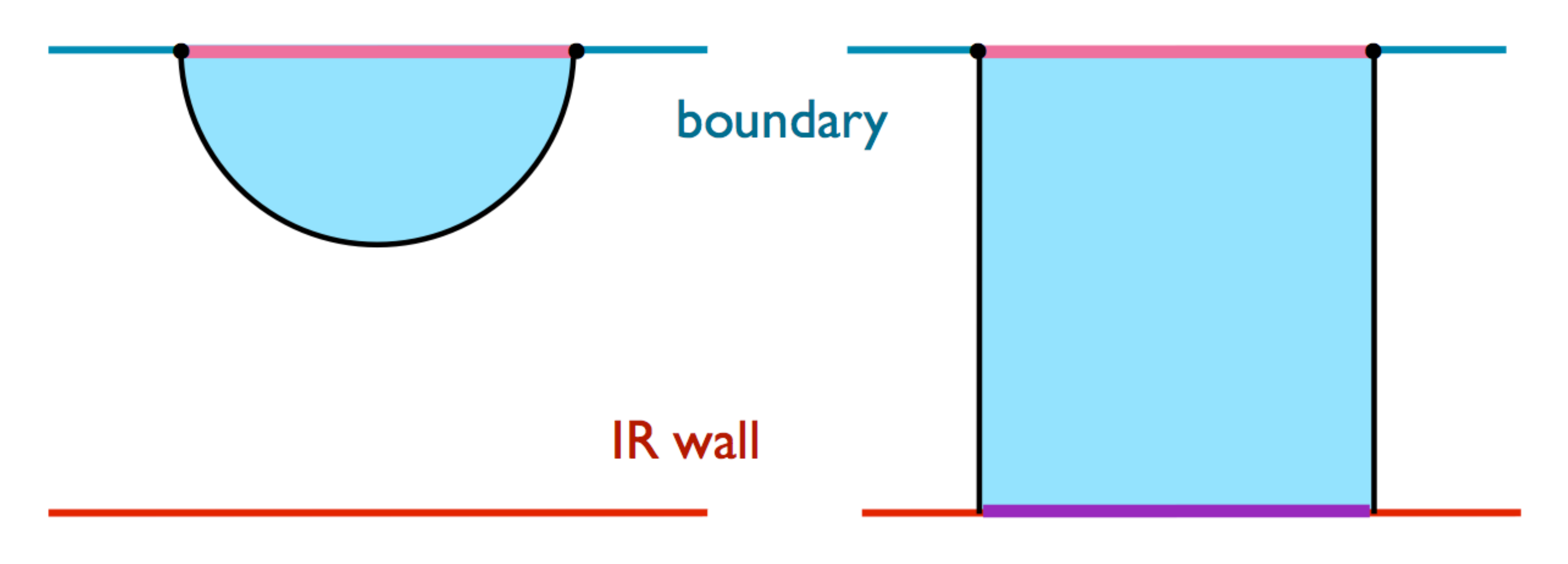}
\end{center}
\vspace{-3mm}
\caption{\label{fig:graph} Connected and disconnected geodesic configurations (black) anchoring at the endpoints of a boundary interval (pink). The interval, the geodesic and possibly an interval at the wall (purple), determine a closed area in the bulk (blue).}
\end{figure}

Since any point in the AdS boundary can be joined by a geodesic with an arbitrary point in the wall,  there is an infinite set of geodesic configurations associated to a given field theory interval which reach the wall. Among them we should select that extremizing the length functional with respect to the endpoints at the wall. When there is more than one geodesic configuration satisfying all the stated requirements, that with smallest length will determine the entanglement entropy \cite{Ryu:2006bv,Ryu:2006ef}.
This is the case for sufficiently small intervals. There is a connected geodesic joining its endpoints without reaching the wall, along with the disconnected configuration composed of two geodesic segments extending from the interval endpoints to the wall. 

\begin{figure}[h]
\begin{center}
\includegraphics[width=5cm]{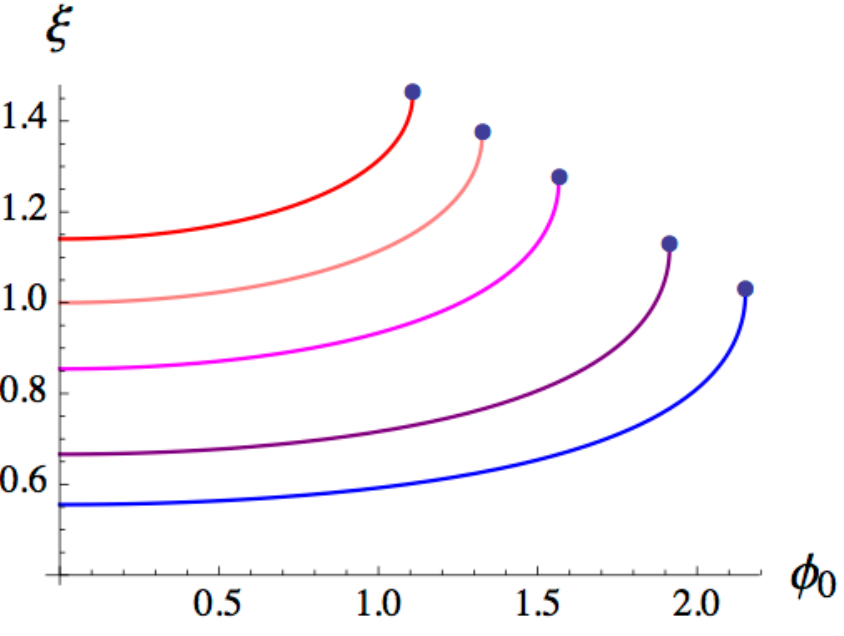}~~~
\includegraphics[width=5cm]{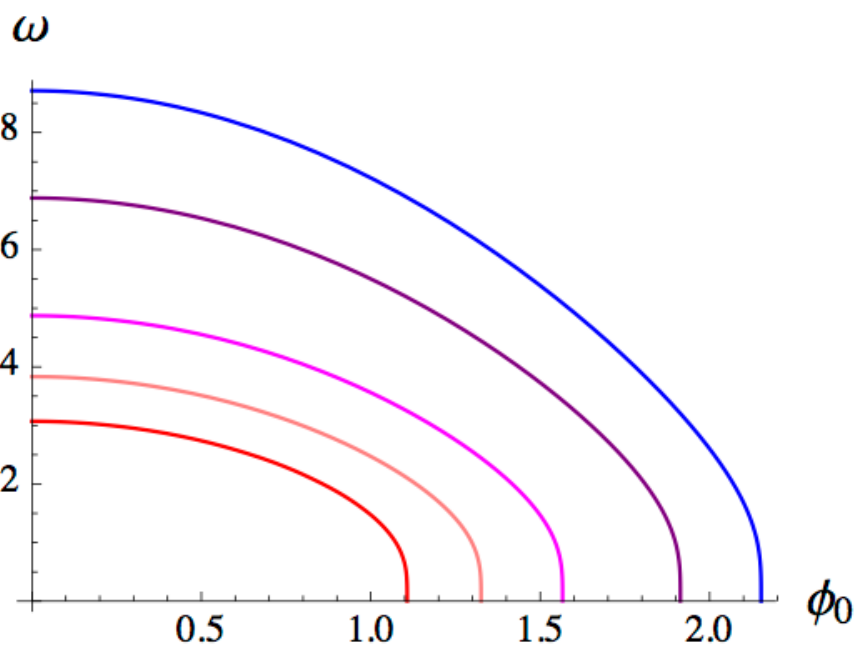}~~~
\includegraphics[width=5cm]{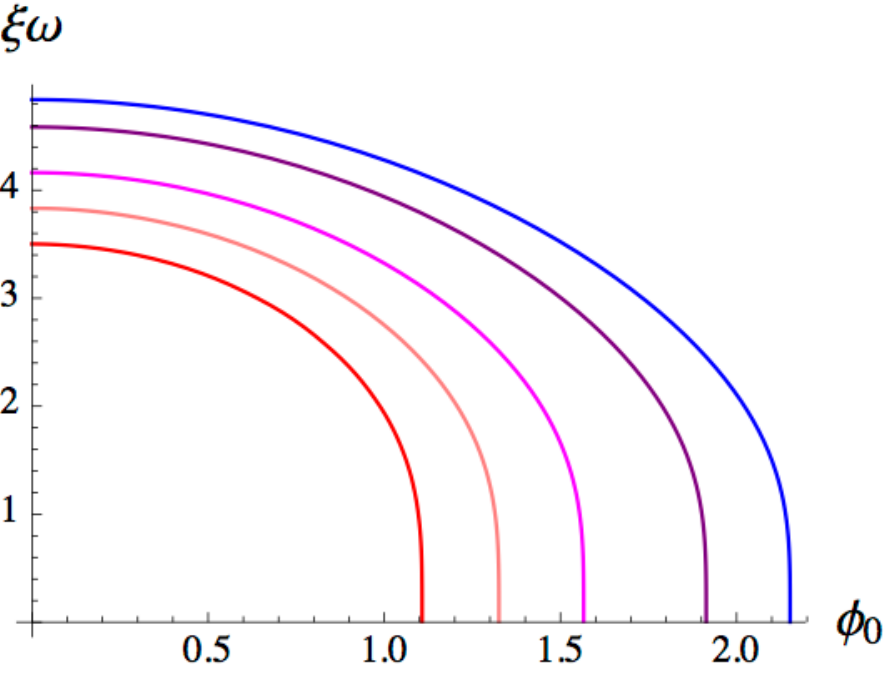}
\end{center}
\caption{\label{fig:lcor} Correlation length (a), mass gap (b) and their product (c) as a function of $\phi_0$ for $M_0\!=\!1/2,0,-1,-4,-8$ from red to blue.}
\end{figure}

Let us consider static solutions of the hard wall model with vanishing scalar profile for simplicity. A geodesic anchoring at the endpoints of an interval of size $l$, as far as it does not reach the wall, is not affected by its presence. The length of such geodesic in a black hole background of mass $M_0$ is
\be
 2\log \left( 2  \sinh \sqrt{M_0} \, l/2 \right) - 2\log \sqrt{M_0} \epsilon\; ,
\label{con}
\ee
where the bulk has been cut at a UV scale $z\!=\!\epsilon$ in order to regularize the otherwise divergent length. The disconnected choice which extremizes the length is given by geodesics starting on the interval endpoints and extending radially to the wall. Its length is
\be
2\log \big(2 z_0/\epsilon \big) - 2\log \Big(  1 + \sqrt{1-M_0 z_0^2} \, \Big) \, .
\label{dis}
\ee
For small intervals the connected choice is preferred, but for large ones it is the disconnected configuration which has smaller length. Once the latter dominates, the entanglement entropy saturates to a value independent of the length of the interval. This is the behavior expected in a gapped theory, whose holographic derivation was sketched in \cite{Ryu:2006ef}. The length for which \eqref{con} and \eqref{dis} coincide gives the correlation length as defined above
\be
\xi= {2 \over \sqrt{M_0}} \, {\rm arcsinh} \left( {\sqrt{M_0} z_0 \over 1+ \sqrt{1-M_0 z_0^2}} \right)  \, .
\label{cor}
\ee 
For vanishing wall mass we obtain, $\xi\!=\!z_0$. The correlation length increases with the wall mass up to a finite value attained at the horizon threshold $M_0 z_0^2\!=\!1$.
Expression \eqref{cor} can be smoothly continued to negative $M_0$, causing a monotonic decrease of the correlation length. For very large negative masses the correlation length tends to zero as
\be
\xi\approx {\pi \over \sqrt{|M_0|}} \, .
\label{asymxi}
\ee
In Fig.\ref{fig:lcor}a we plot $\xi$ as a function of $\phi_0$ for several fixed values of the wall mass. 

At the topological level, all values $M_0 z_0^2\!<\!1$ give equivalent geometries.  In particular, the entanglement entropy of an interval is equivalent to that of its complementary region for all of them. It follows immediately from requiring that  a region at the AdS boundary and its associated geodesic are homologous in the relative homology defined with respect to the infrared wall. This suggests that all geometries with $M_0 z_0^2\!<\!1$ describe pure states \cite{AbajoArrastia:2010yt,Takayanagi:2010wp}.

An important information on the static backgrounds is their spectrum of normal modes. The holographic dictionary relates it with the non-analyticities in the two-point function of the dual field theory operator $\cal O$. Hence we will use the lowest normal frequency as definition of the mass gap, which we plot in Fig.\ref{fig:lcor}b. Both the correlation length and the mass gap give a measure of the characteristic infrared scale of the theory. In an interacting theory we do not expect both quantities to coincide, but they should be of the same order. Fig.\ref{fig:lcor}c shows that this is verified, except close to the stability threshold, where $\omega$ vanishes while $\xi$ remains finite. The correlation length has been calculated using the geodesic length, which gives the leading N contribution. However the mode becoming unstable close to the threshold for horizon formation reflects an order 1 effect. For this reason $\xi$ and $\omega$ measure different aspects of the physics in that limit, and thus give independent answers. For very large negative wall mass and vanishing scalar profile, where the correlation length is given by \eqref{asymxi}, we have obtained numerically
\be
\xi \omega \approx 6.5 \, ,
\ee
in accord as well with the field theory reasoning.

A necessary consistency check on the backgrounds with negative wall mass is that they do not induce superluminal propagation at the boundary. Indeed, we have evaluated the dispersion relation for several normal modes on static backgrounds with negative $M_0$, and always found group velocities smaller than one. A related check requires that, after reflection on the hard wall, light rays leaving the AdS boundary return to it within the light-cone of the starting point. For static solutions with a vanishing scalar profile, the distance travelled by such light ray verifies \footnote{Null geodesics are described by
\be
z \sqrt{M_0}= \sinh \gamma \sinh (\sqrt{M_0} x) \, , \hspace{1cm}  t \sqrt{M_0}=  {\rm arctanh} \big(\! \cosh \gamma \tanh (\sqrt{M_0} x) \big) \, ,
\nonumber
\ee
which ca be analytically continued to negative masses.}
\be
\tanh {\sqrt{M_0}\,  \Delta t \over 2} = \cosh \gamma  \, \tanh{ \sqrt{M_0} \, \Delta x \over 2} 
\label{light}
\ee
where $\gamma$ determines its incidence angle with the AdS boundary: ${dz \over dx}\big|_{z\!=\!0}\!=\! \sinh \gamma$. Hence $\Delta t \!\geq \! \Delta x$. Relation \eqref{light} extends with the same conclusion to backgrounds with negative wall mass as far as $\sinh  \gamma \! \geq \! \sqrt{|M_0|} z_0$. Smaller incidence angles give rise to null geodesics that get repelled towards the AdS boundary without reaching the wall. Independently of $\gamma$ they verify
\be
\Delta t =\Delta x = {\pi \over \sqrt{|M_0|}} \, .
\ee
They return to the boundary on the light cone of the starting point, respecting thus causality.

\section{Time-dependent boundary conditions at the wall}

\begin{figure}[h]
\begin{center}
\includegraphics[width=6.8cm]{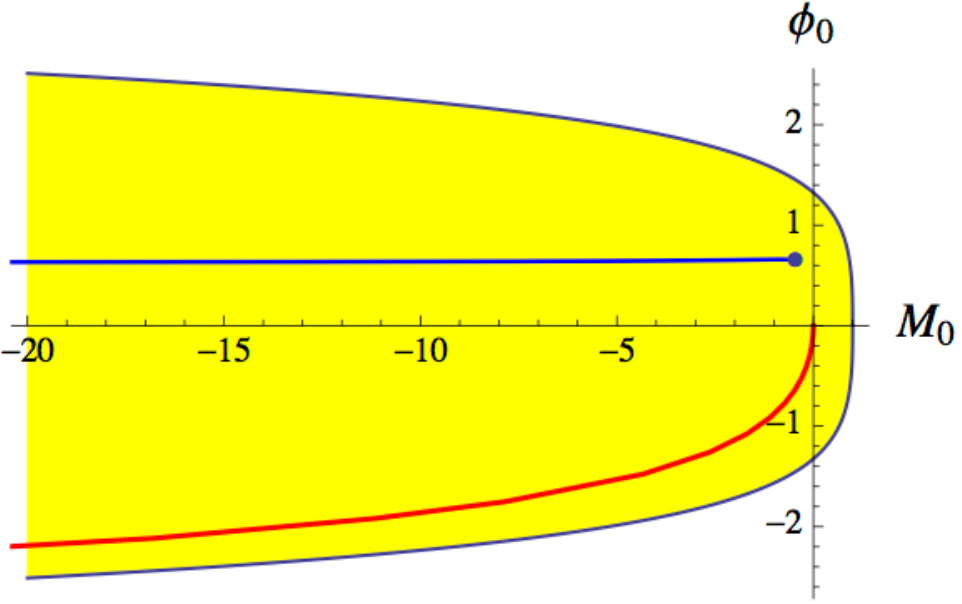}~~~~~
\includegraphics[width=6cm]{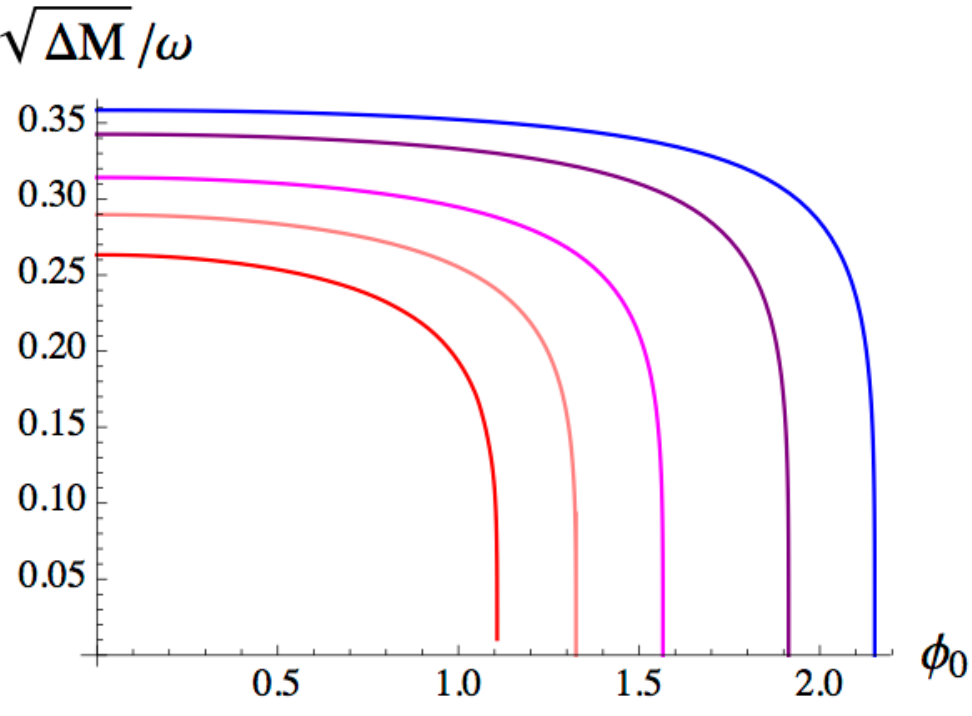}
\end{center}
\caption{\label{fig:phase} Left: In yellow, couplings which admit stable solitonic solutions. Final couplings after an action \eqref{bcwall} which starts at $M_0\!=\!\phi_0\!=\!0$: as a function of $\alpha$ for $\epsilon\!=\!0.3$ (blue), and as a function of negative $\epsilon$ for $\alpha\!=\!.8$ (red). Right: Relation \eqref{testHW} as a function of $\phi_0$ for fixed $M_0$, with the same color code as in Fig.\ref{fig:lcor}.}
\end{figure}

We have seen that the infrared physics is not only determined by the position of the wall. The correlation length and the mass gap depend strongly on $M_0$ and $\phi_0$. Since boundary conditions can not be changed by bulk dynamics these parameters should be interpreted as couplings, in accordance with the usual holographic dictionary. 
Parameters in the yellow area of Fig.\ref{fig:phase} admit stable static solitonic configurations. These configurations provide the ground states of the dual field theory for different couplings.
Only pure AdS with a cutoff, associated to $M_0\!=\!\phi_0\!=\!0$, respects Lorentz invariance. Other backgrounds represent pure states with order $N^2$ expectation values of the stress tensor and the operator $\cal O$ dual to the scalar field. Ground states of these characteristics describe condensates. Therefore generic couplings $\{M_0,\phi_0\}$ create a non-trivial medium in the dual field theory.

We can simulate an action on the dual field theory in which both $\{M_0,\phi_0\}$ vary, by imposing time dependent boundary conditions at the wall. 
An static state is assumed at early times. We consider boundary conditions of the form \eqref{bc} with the particular profile
\be
\Pi_0(t)={\epsilon \over \alpha}{1  \over \cosh^2 {t \over \alpha}} \; .
\label{bcwall}
\ee 
which forces the value of $\phi_0$ to change. In the process a scalar pulse is generated which enters from the wall into the bulk. As we explained in Section 2, the total mass \eqref{mass} is conserved under generic boundary conditions at the wall. Hence the scalar pulse draws its energy from the wall, determining in this way the final value of $M_0$. In consequence $M_0$ can be rendered negative as a result of \eqref{bcwall}. This is always the case when the initial background is pure AdS with an infrared cutoff. 

Geometries with $M_0\!<\!0$ lead to regular evolutions of the homogeneous Einstein equations much as non-negative values do, providing one more check of their consistency in the context of the hard wall model.
The blue line in Fig.\ref{fig:phase} shows the final couplings reached after an action \eqref{bcwall} starting from pure AdS, as a function of $\alpha$ at fixed $\epsilon\!=\!0.3$. Slow actions, namely $\alpha\!\rightarrow\!\infty$, drive the system to the point signaled with a dot. Decreasing $\alpha$ produces scalar pulses sharply localized on the radial direction of increasingly large mass, which in turn implies large negative $M_0$.
The value of $\phi_0$ keeps practically constant, showing that it is mainly determined by $\epsilon$. For completeness, we plot in red the final couplings as a function of $\epsilon$ at fixed $\alpha\!=\!.8$.

The Schwarzchild radius for most solitonic solutions is below the wall position, see Fig.\ref{fig:static}a. Conservation of energy forbids that actions \eqref{bcwall} starting on such states lead to the formation of a horizon. Instead they will result in a scalar pulse that bounces forever back and forth between AdS boundary and infrared wall \cite{Craps:2013iaa}. Gravitational collapse ending in the formation of a black hole holographically models quantum field theory thermalization. On the contrary, oscillating geometries where interpreted in \cite{Abajo-Arrastia:2014fma,daSilva:2014zva} as geometrical counterparts of quantum revivals in the unitary evolution of the dual field theory. Quantum revivals denote periodic reconstructions of an initial out of equilibrium state \cite{Robinett2004}. Interacting field theories typically exhibit a fast approach to a thermal state, which prevents revivals. Hence oscillating geometries should be disfavored. A natural exception to this rule can be found in out of equilibrium states where the scale set by the energy density is smaller than the infrared gap. Namely, revivals in d-dimensional field theory might happen only when
\be
\big(E/n)^{1 \over d}  < m_{gap} \, ,
\label{testR}
\ee
where $m_{gap}$ is a measure of the gap, $E$ the energy density and $n$ the number of elementary fields. The dynamics of infalling matter shells in global AdS was found to be consistent with this relation \cite{Abajo-Arrastia:2014fma,daSilva:2014zva}. In that case $m_{gap}$ is set by the inverse of the boundary sphere radius and $n\!\sim\!c$, the central charge of the dual CFT. 

A first check in order to interpret oscillating geometries in the hard wall model as dual to quantum revivals is hence \eqref{testR}. For field theories which admit a holographic representation, $n\!\sim\! 1/G$, with $G$ the Newton's constant. Notice that in Einstein's equations \eqref{eqforphi}-\eqref{eqforA} we have eliminated the factor $8 \pi G$ which multiplies the matter energy momentum tensor by a convenient rescaling of the scalar field. Due to that $M$, the total mass as defined in \eqref{mass}, represents the energy density per species of the dual field theory. 

Since the boundary conditions determine the concrete dual field theory, we focus on scalar configurations with fixed $\{M_0,\phi_0\}$. 
Among this set, we define $M^{th}$ by the condition that any configuration with $M\!>\!M^{th}$ leads to horizon formation by direct collapse. 
An estimate for this threshold consists in requiring the associated Schwarzchild radius to lie at the wall, $M^{th} \!=\!z_0^{-2}$.
We have seen however that there are stable solitonic solutions whose Schwarzchild radius lies above the wall. Thus a better estimate is provided by
\be
M^{th} \lesssim {\overline M} \, ,
\ee
with ${\overline M}$ the mass of the limiting solitonic solution for the chosen $M_0$, in the border of the yellow area in Fig.\ref{fig:phase}a. Relation \eqref{testR} translates then into
\be
\sqrt{\, {\overline M}-M_{gs}} \, \lesssim \, m_{gap} \, ,
\label{testHW}
\ee
where $M_{gs}$ and $m_{gap}$ refer to the solitonic solution with boundary conditions $\{M_0,\phi_0\}$.
The quantity inside the square root is the minimal energy density per species over the ground state which will always lead to fast thermalization. Relation \eqref{testHW} states that the scale it sets should not be parametrically larger than the infrared gap. Choosing for $m_{gap}$ the frequency of the lowest normal mode, Fig.\ref{fig:phase}b shows that this is satisfied on subspaces of coupling space with constant wall mass. We complement this test by analyzing \eqref{testHW} for asymptotical large negative $M_0$ and vanishing scalar profile. It is now convenient to identify $m_{gap}\!=\!1/\xi$. We have checked that in the limit of large negative wall masses, the quotient ${\overline M}/|M_0|$ tends to zero. Using this and \eqref{asymxi} we have
\be 
\xi \sqrt{{\overline M}+|M_0|}  \rightarrow \pi \, ,
\ee
which agrees again with \eqref{testHW}.

\section{The holographic dual of a quench}

We have just seen that the oscillating geometries of the hard wall model satisfy the basic requirement to be interpreted in terms of quantum revivals. In the following we want to explore what is the field theory state associated with an oscillating geometry and the physical mechanism for the revivals it represents. 

The time parameter $\alpha$ has a major influence on the properties of the scalar pulse resulting from \eqref{bcwall}.
We have chosen $t$ to be the proper time at the boundary. Hence 
$\alpha$ should be corrected by a redshift factor
\be
\tilde \alpha={\alpha \over z_0}  \sqrt{A_0} \, e^{-\delta_0}\, ,
\ee
where we have used \eqref{metric}. The metric functions $\delta$ and $A$ have been evaluated at the wall, on the solitonic configuration chosen as initial state.
Values $\tilde \alpha\!<\!1$ result in a radially localized pulse that travels between boundary and wall. Large values lead to adiabatic evolutions where the system moves along the space of solitonic solutions without generating a dynamical profile. Actions for which $\tilde \alpha$ is of the order one, mostly excite the lowest normal mode over the soliton determined by the final boundary conditions at the wall. The oscillating component of the scalar profile is in this case a standing wave. 

\begin{figure}[h]
\begin{center}
\includegraphics[width=5cm]{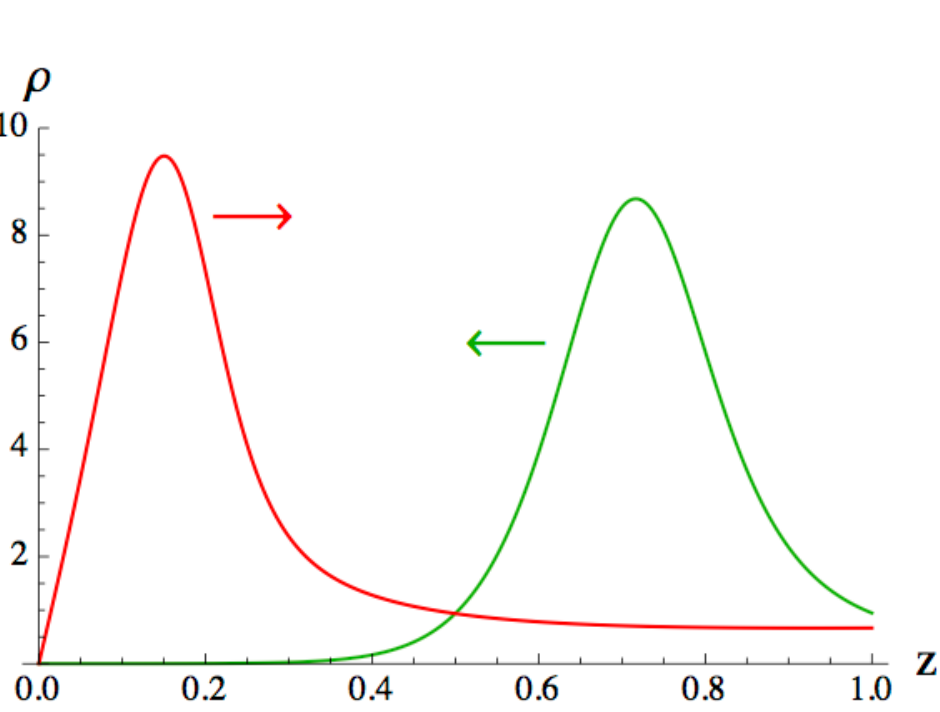}~~~
\includegraphics[width=5cm]{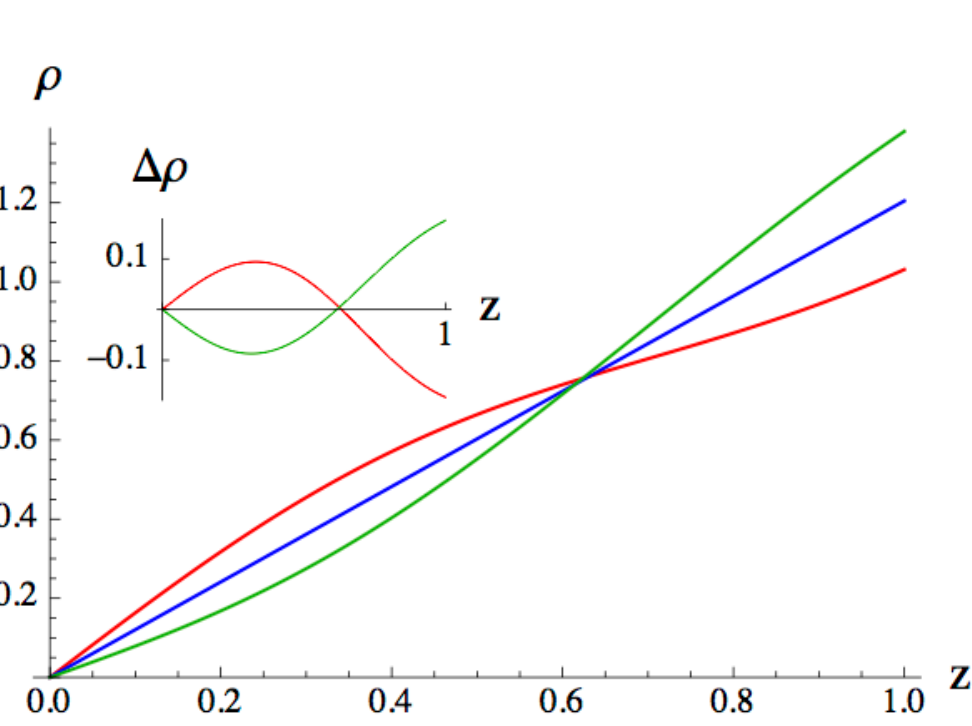}~~~
\includegraphics[width=5cm]{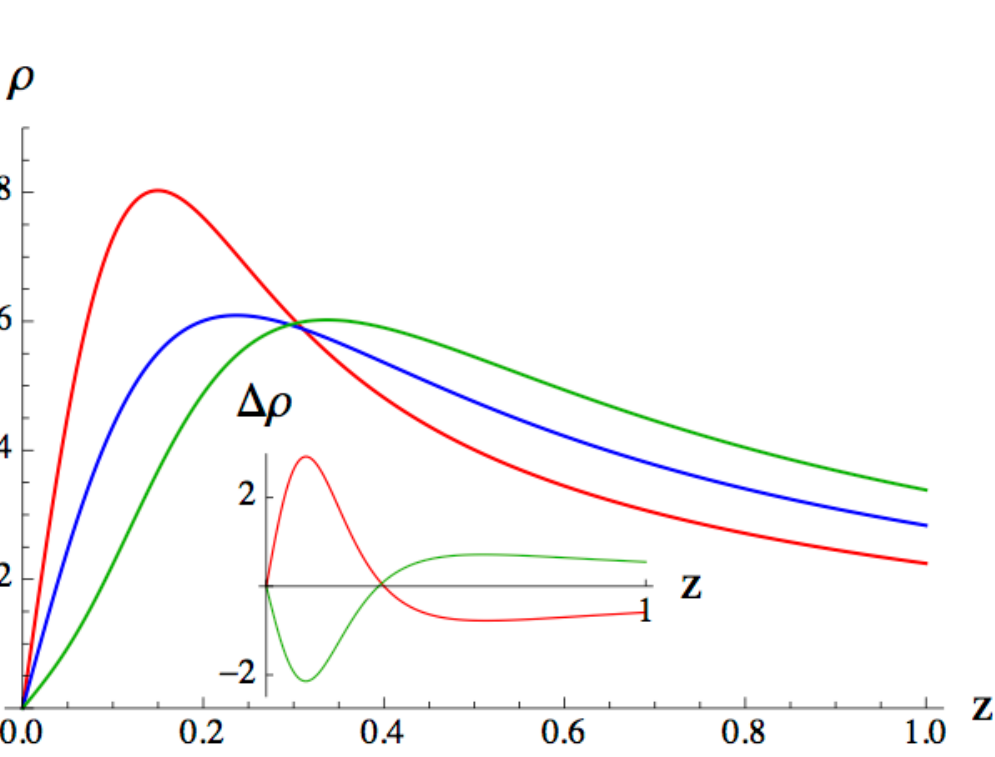}
\end{center}
\caption{\label{fig:profile} Scalar pulses resulting from \eqref{bcwall} with initial state $\{M_0,\phi_0\!=\!0\}$: a) $\alpha\!=\!.2$, $M_0\!=\!0$; b) $\alpha\!=\!.8$, $M_0\!=\!0$; c) $\alpha\!=\!.2$, $M_0\!=\!-18$. The final values of the couplings are: a) $M_0\!=\!-1.15$, $\phi_0\!=\!0.66$, b) $M_0\!=\!-0.3$, $\phi_0\!=\!0.52$, c) $M_0\!=\!-20.17$, $\phi_0\!=\!0.58$. Inset: scalar pulse after subtracting the soliton profile associated to the final couplings.}
\end{figure}

This behavior is illustrated in Fig.\ref{fig:profile} for several examples with $z_0\!=\!1$. The initial state for Fig.\ref{fig:profile}a and \ref{fig:profile}b is the Lorentz invariant vacuum, for which ${\tilde \alpha}\!=\!\alpha$. The first action has $\alpha\!=\!0.2$ and generates a traveling scalar pulse, from which we have taken two snapshots. The second has a longer time span, $\alpha\!=\!0.8$, and induces a standing wave over the corresponding solitonic state, depicted in blue. This is better appreciated in the inset, where the contribution of the static soliton has been subtracted. The action in Fig.\ref{fig:profile}c starts from $M_0\!=\!-18$ with vanishing scalar profile, and has $\alpha\!=\!0.2$ but ${\tilde \alpha}\!=\!0.87$. Consistently it results in a standing wave again. 

Following the analysis in the end of the previous section a measure of the energy density per species in the associated field theory state, normalized with respect to the infrared gap, is provided by 
\be
e={M-M_{gs} \over {\overline M} - M_{gs}} \, .
\label{quotient}
\ee
$M$ is the total mass, which remains constant along the wall action, $M_{gs}$ the mass of the static soliton associated to the final value of the couplings, and $\overline M$ as in \eqref{testHW}.
This quotient is small for actions \eqref{bcwall} with ${\tilde \alpha}\!\approx\!1$, indicating that the dual field theory states are in the limit of small energy density. Indeed, the examples of Fig.\ref{fig:profile}b,c have 
$0.001$ and $0.002$ respectively. Actions with $\tilde \alpha$ small lead in general to higher excited states. The pulse in Fig.\ref{fig:profile}a is generated by an action with ${\tilde \alpha}=\!0.2$ and the energy density as measured by quotient above takes the value $0.28$. 

\begin{figure}[h]
\begin{center}
\includegraphics[width=6.5cm]{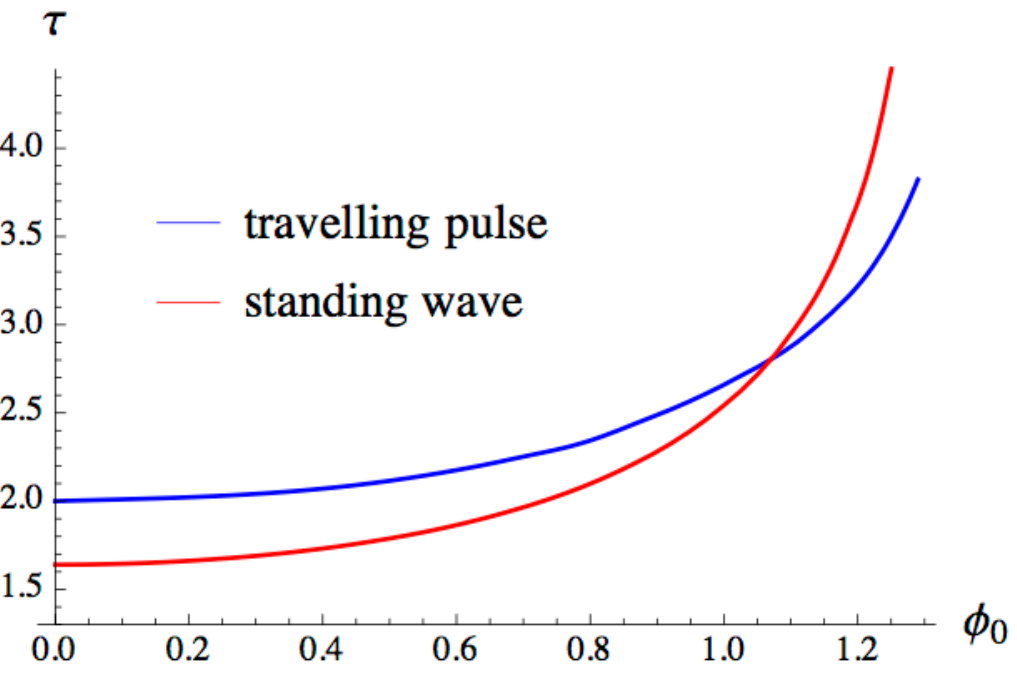}~~~
\includegraphics[width=8cm]{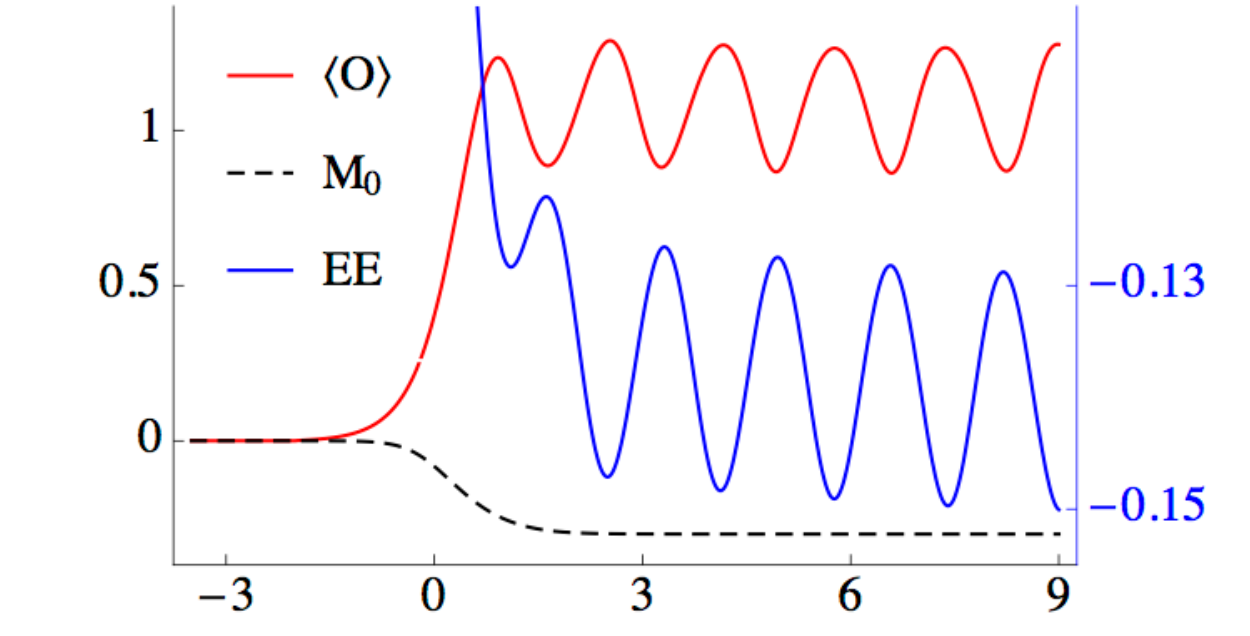}
\end{center}
\caption{\label{fig:period} Left: Periodicity of oscillating geometries as a function of $\phi_0$ for $M_0\!=\!0$. Right: Evolution of $\langle {\cal O} \rangle$, $M_0$ and the entanglement entropy of large intervals for the example in Fig.\ref{fig:profile}b.}
\end{figure}

A relevant information about oscillating geometries is their periodicity. This uncovers an important difference between geometries based on traveling versus standing pulses.
The period of standing configurations is determined by the frequency of the lowest normal mode
\be
\tau \approx {2 \pi \over \omega} \; .
\label{period1}
\ee
The period of well localized pulses is of the same order, but their numerical values do not coincide. For $\phi_0$ far from the instability threshold, standing pulses oscillate with a neatly smaller periodicity. 
This is shown in Fig.\ref{fig:period}a for configurations with $M_0\!=\!0$. We plot the period of the lowest normal mode and that of thin traveling pulses of low mass. When $\phi_0$ approaches the instability threshold, $\omega$ tends to zero and \eqref{period1} diverges. Since thin pulses have almost no overlap with the lowest harmonic, their period however remains finite in that limit.

We would like to confront the hard wall model with the out of equilibrium dynamics of known QFTs. The requirements needed for a QFT to have a classical gravity dual, strong coupling and large N, make such theories difficult to analyze with ordinary tools. Tensor network techniques \cite{Orus:2013kga} improve on this situation since they are not limited by the strength of the interactions.
These techniques have been applied recently, in particular, to the study of the Schwinger model \cite{Banuls:2013jaa}-$\!\!$\cite{Pichler:2015yqa}. In \cite{Buyens:2013yza} the authors explored the quench dynamics of the massive Schwinger model, triggered by turning on abruptly an uniform background electric field. The theory after the quench remains gapped. Switching on an electric field generically induces pair production of fermions. However when the fermions are massive this requires an energy density of the order of the infrared gap. Below it, the system was found to undergo revivals \cite{Karelnotes}. 

The revivals cause the periodic oscillation of the electric field itself and the particle number. In the limit of small quenches the revivals are almost perfect with periodicity
\be
\tau= {2 \pi \over m} \, ,
\label{revivalsS}
\ee
where $m$ is the mass of the lowest stable excitation of the final hamiltonian. This periodicity allows a guess on the composition of the state created by the quench. It is reasonable to associate it with some coherent state of momentum zero modes of the lowest excitation \cite{Buyens:2013yza}\cite{Karelnotes}. The oscillations in the particle number induce also in phase oscillations of the entanglement entropy \cite{Karelnotes}\cite{Pichler:2015yqa}. 
The Schwinger model of course does not satisfies the holographic requirement of having a large number of elementary fields. On the other hand, holographic revivals mostly depend on the ratio of energy density per species. Since N factors out in this quotient, we believe that a qualitative comparison between the phenomenology of revivals in the a priori very different Schwinger and hard wall models can be sensible.

Recall that holography relates the frequencies of the normal modes with the location of non-analyticities in correlation functions. Therefore it is natural to relate the mass of the lowest stable excitation in the Schwinger model with the frequency of the lowest harmonic in the hard wall model. Due to this, although the Schwinger model undergoes a sudden quench, the periodicity \eqref{revivalsS} of its revivals strongly suggests their holographic representation in terms of a standing pulse. 
Notably this is also consistent with a dual field theory state composed mostly of excitations with vanishing momentum. It has been argued that the radial infall of a collapsing mass profile holographically represents the drift away of entanglement excitations after a quench \cite{Abajo-Arrastia:2014fma}\cite{daSilva:2014zva}. However when the field theory is in a state dominated by zero momentum modes, propagation effects should not play a role. This again rules out radially localized traveling pulses. Instead no notion of radial displacement can be associated to standing waves. In spite of this, they induce oscillations in the entanglement entropy. In Fig.\ref{fig:period}b we have plotted the time evolution of the wall action in Fig.\ref{fig:profile}b. We observe the oscillatory behavior of the expectation value of the operator dual to the scalar bulk field (red) and the entanglement entropy (blue). The oscillations of $\langle {\cal O} \rangle$ can be put into correspondence with those of the electric field, which in the Schwinger model are in opposite phase to those of the particle number and the entanglement entropy.

\begin{figure}[h]
\begin{center}
\includegraphics[width=6.5cm]{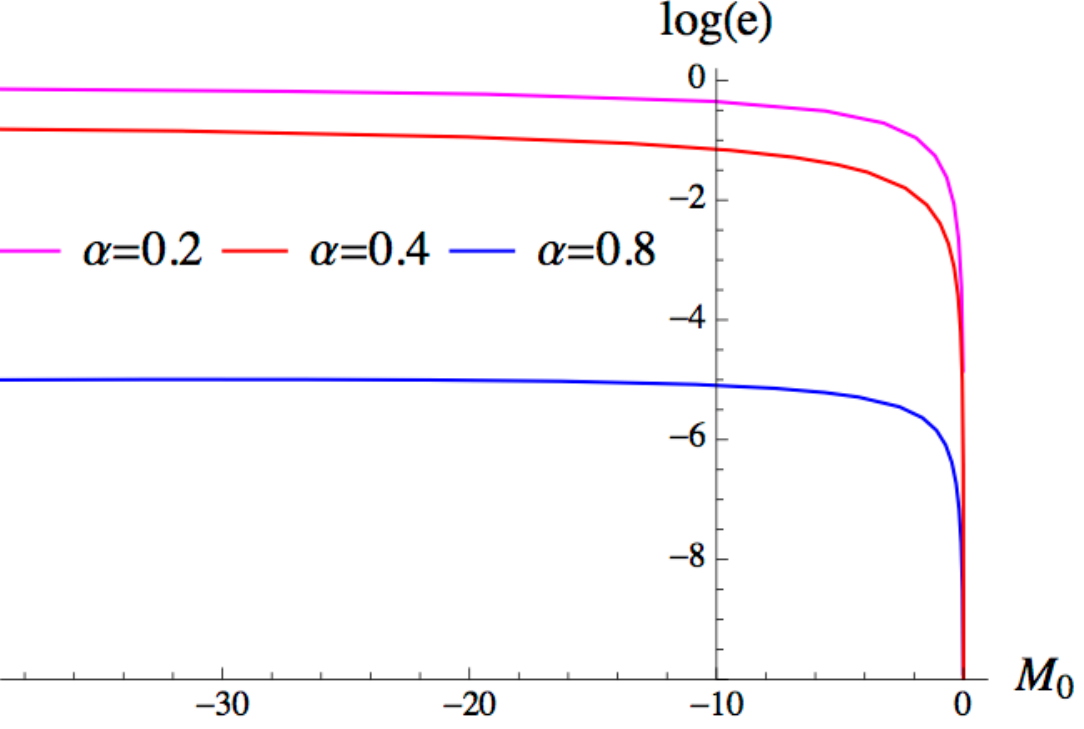}~~~
\includegraphics[width=6.5cm]{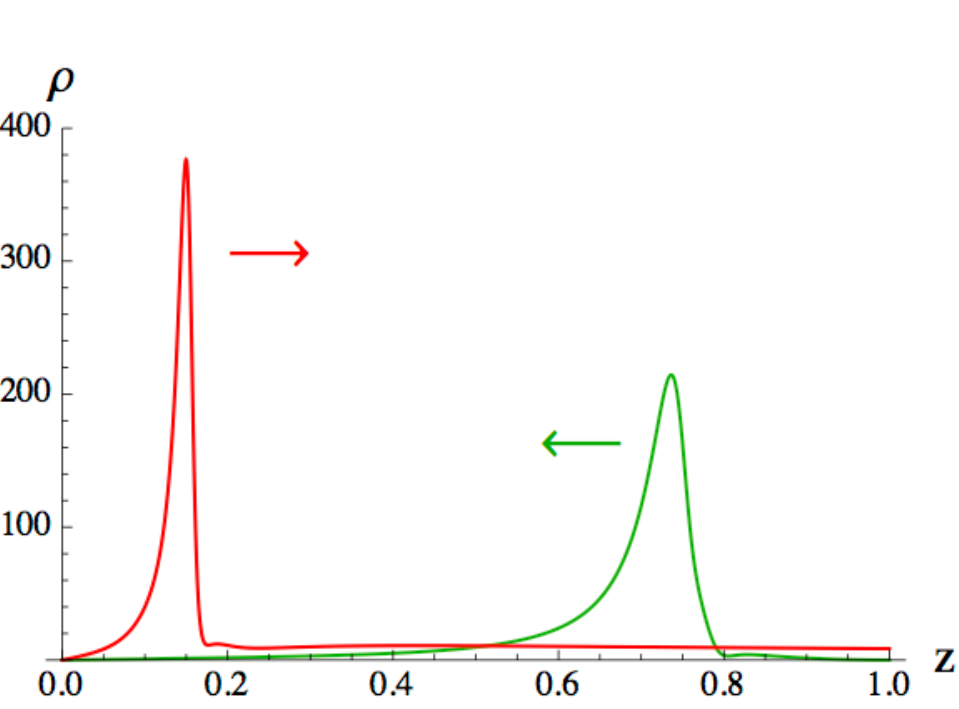}
\end{center}
\caption{\label{fig:travel} Left: Value of the quotient \eqref{quotient} for wall actions with initial state $\{M_0\!=\!0,\phi_0\!=\!0\}$ and fixed $\alpha$, as a function of the final wall mass. Right: As in Fig.\ref{fig:profile}a, with final couplings $M_0\!=\!-10.09$, $\phi_0\!=\!1.27$.}
\end{figure}

Let us move to quenches with a higher energy density, but still compatible with revivals. The resulting out of equilibrium state can be expected to include, besides momentum zero, an appreciable percentage of finite momentum modes. Hence propagation effects will be relevant in the evolution of its entanglement pattern. Since propagation of entanglement has been holographically linked to radial displacement in the bulk, a geometry modeling them should not be based on a perfect standing wave. The hard wall model is also consistent this expectation. We have mentioned that standing waves resulting from wall actions have ${\tilde \alpha}\!\approx\!1$ and very small energy density. Larger values can be only obtained for ${\tilde \alpha}\!<\!1$. And this implies some degree of radial displacement on the induced pulse, as can be seen in Fig.\ref{fig:profile}a. 
A detailed illustration of the relation between $\alpha$ and the resulting energy density is provided in Fig.\ref{fig:travel}a, by evaluating \eqref{quotient} for wall actions starting in the Lorentz invariant vacuum with fixed $\alpha$ and different final $M_0$. We plot $\log e$ in order to compare the results for different $\alpha$ in the same graphic. We observe that its value does not vary much along a wide range of final negative wall masses, but strongly depends on $\alpha$. 
Standing waves with $e\!<\!0.01$ are always obtained for $\alpha\!=\!0.8$. Large negative masses give $e\!\approx\!0.8$ for $\alpha\!=\!0.2$, close to the threshold for gravitational collapse or equivalently fast thermalization. The radial localization of the scalar profile resulting from a wall action neatly increases with $e$ even at fixed $\alpha$. Compare to this aim the $\alpha\!=\!0.2$ examples in Fig.\ref{fig:profile}a, with $e\!=\!0.28$, and Fig.\ref{fig:travel}b, with $e\!=\!0.7$.

As an additional test we evolve an initial profile based on the lowest normal mode with finite amplitude
\be
\Phi(z,t\!=\!0)=\Phi_{S}(z) + \epsilon \,  {d\chi \over dz}(z) \, , \hspace{1cm}  \Pi(z,t\!=\!0)=0 \, ,
\label{idata}
\ee
keeping the wall couplings fixed. The function $\Phi_S$ is the soliton solution associated to them, and $\chi$ the radial profile of its lowest normal mode
\be
\Delta \phi (z,t)=\chi(z) \cos(\omega t) \, .
\label{harmonic}
\ee
These initial data represent a field theory coherent state of zero modes of the lowest excitation with an average occupation number controlled by $\epsilon$. The resulting radial dynamics for $M_0\!=\!-1$, $\phi_0\!=\!0.5$ and $\epsilon$ tunned to obtain $e\!=\!0.5$ is shown in Fig.\ref{fig:narrow}a. We have plotted the radial profile minus its static component, as in the inset of Fig.\ref{fig:period}b,c. 
By comparing with the initial energy distribution, brown dashed curve, we observe the emergence of localized structures.
We have checked that radial localization and displacement is more pronounced the bigger the value of $e$, in agreement with the field theory interpretation.

\section{Discussion}

In this final section we want to address at a more general level how to holographically model a quench protocol. The holographic dictionary maps the boundary value of bulk fields to field theory couplings. We have shown that this applies at the infrared wall as well as at AdS boundary. Time dependent boundary values provide therefore the natural playground to represent quantum quenches.

\begin{figure}[h]
\begin{center}
\includegraphics[width=6.5cm]{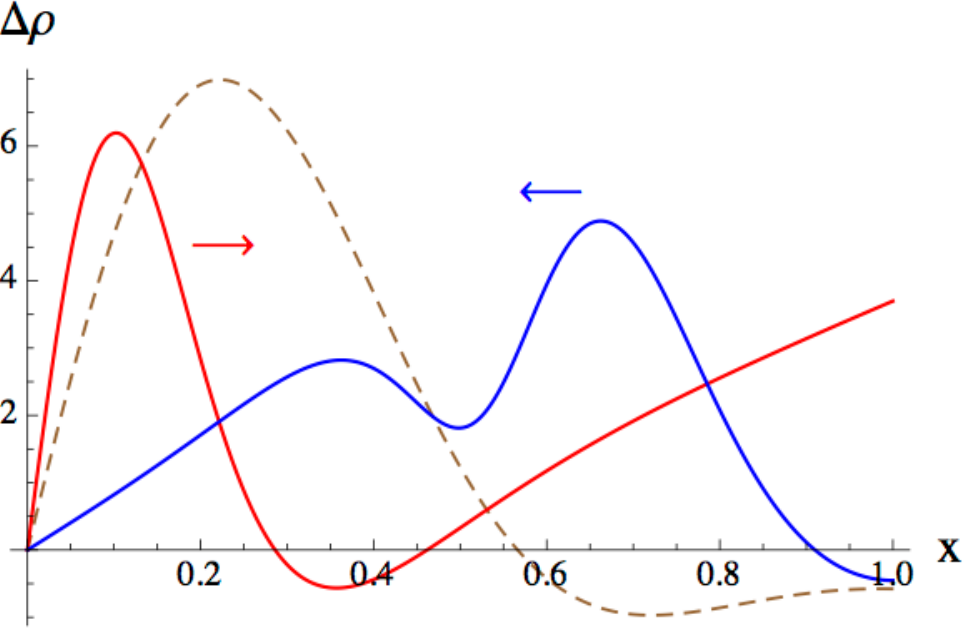}~~~
\includegraphics[width=7.2cm]{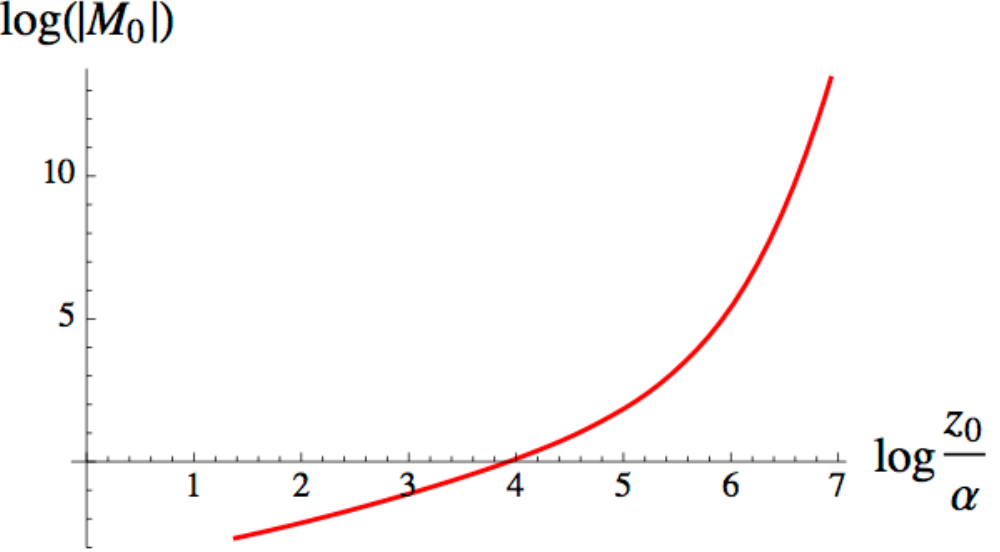}
\end{center}
\caption{\label{fig:narrow} Left: Scalar profile minus its static component for the initial data \eqref{idata} with $M_0\!=\!-1$, $\phi_0\!=\!0.5$ and $e\!=\! 0.5$. We show two snapshots of the evolution and, with a dashed line, the initial profile. Right: Final value of the wall mass for increasingly abrupt wall profiles of fixed amplitude, starting at $\{M_0\!=\!0,\phi_0\!=\!0\}$.}
\end{figure}

A quench is understood as a change of couplings faster than any response time proper of the system. For this reason a quench is usually described as an instantaneous action.
A first guess is identifying the time span for the variation of the boundary conditions with the time span of the quench. This is however problematic  \cite{Buchel:2013gba}-$\!\!$\cite{Das:2015jka}.
By dimensional arguments, varying the boundary value of a massless scalar field at the AdS$_3$ boundary generates a pulse with mass and width
\be
M \propto {\epsilon^2 \over \delta t^2} \, , \hspace{1cm}  \Delta \propto \delta t \, ,
\label{bcAdS}
\ee
where $\delta t$ is the time span of the change and $\epsilon$ its amplitude. We wish to take the limit $\delta t\!\rightarrow\!0$ keeping $\epsilon$ non vanishing, since otherwise no change of couplings is generated. A singular shell is then obtained carrying infinite energy and zero width. A related situation is obtained for the wall actions \eqref{bcwall}. Decreasing $\alpha$ leads to ever more negative wall masses. In Fig.\ref{fig:narrow} we plot the final value of $|M_0|$ when the wall action starts in the Lorentz invariant vacuum. Its growth with $1/\alpha$ is much faster than a power law. Notice that is possible because two dimensionfull quantities are now involved, $\alpha$ and $z_0$.
Relating the time span of a quench with $\alpha$ would lead again to a singular configuration.

We are interested in holographically simulating global quenches which induce a finite energy density. In order to achieve this in the context of gravity plus a massless scalar, we are forced to consider changes of boundary conditions with a finite time span. By comparing with the massive Schwinger model  \cite{Buyens:2013yza}, we have seen that this crude modeling leads to reasonable results. A crucial condition was linking the radial thickness of the resulting scalar configuration, with the energy density of the quench state we want to describe. We expect this condition to be an important entry in the dictionary for holographic quenches. It is supported by two facts that go beyond the particular model considered in this paper. Namely, the different radial dynamics of broad and thin profiles, and the fact that increasing the mass of a broad profile leads to the appearance of localized structures as it evolves.

A closely related question is what type of bulk configurations can not be expected to represent typical out of equilibrium field theory states. By typical we mean that no fine tuning was necessary to generate them.
This is especially relevant in an initial data setup for the gravitational evolution, as in \eqref{idata}. Indeed the dual interpretation of initial pulses of arbitrary shape is still an open issue.
We argue now that thin pulses of small mass, or more precisely small $e$ \eqref{quotient}, are not typical in the above sense.

In the context of the hard wall model, let us consider a thin pulse whose mass is small enough for the associated Schwarzchild radius to be well below $z_0$.
Following previous arguments, a thin pulse must correspond to a dual field theory state able to resolve scales much shorter than the correlation length. 
On the other hand, the assumed mass implies that there is not
enough energy density to distribute it democratically between the $N^2$ field theory degrees of freedom and overcome, or even come close to the gap. In the language of the Schwinger model, this suggests the production of virtual pairs which do not produce an irreversible decay of the electric field. Virtual pairs can not propagate further than the correlation length before recombining, what is consistent with the picture of a thin pulse traveling  back and forth between wall and AdS boundary. It is unlikely that such a field theory state, which combines characteristics proper of high and low energies, can be produced without fine tuning. 

In spite of these issues, thin pulses of small mass were important for interpreting oscillating geometries in global AdS in terms of quantum revivals \cite{Abajo-Arrastia:2014fma,daSilva:2014zva}. Their periodicity  is
\be 
\tau \approx \pi R \, ,
\label{tauglobal}
\ee
with $R$ the radius of the global AdS boundary sphere, where the dual field theory lives.
Revivals of a free theory on a finite size space have the same periodicity, which also agrees with  the simple quasiparticle picture of \cite{Calabrese:2005in}.
Broad pulses have a different period, inherited from the lowest normal mode and difficult to interpret in purely field theory terms. 
An additional hint for the relation of thin pulses with field theory states of high energy is the following.
The evolution of rational CFT's on a circle after a quench with energy much larger than the inverse radius was studied in \cite{Cardy:2014rqa}. As it happens for holographic models based in thin pulses, the overlap of the initial state with the state describing the system at time $t$ experiences a very fast decay. Subsequently the initial state is partially reconstructed with periodicity \eqref{tauglobal}. 
It is the very constrained dynamics of RCFT's what allows to overcome the energy bound for revivals \eqref{testR}, which holographic theories respect \cite{deBoer:2016bov}. 

Finally we want to stress some properties of the very simple hard wall setup considered in this paper. 
By introducing time dependent boundary conditions at the wall, we move in coupling space between theories with different infrared physics. The wall acts as an infinite reservoir from which to drag energy at the price of increasing the mass gap of the dual theory. Hence we come closer to a quench scenario than just simulating an energy injection without real change in the hamiltonian. 
Moreover, the energy density resulting from a wall action \eqref{bcwall} is mainly determined by one of its two parameters, the time span.
This renders again more direct the map with a quench. These characteristics make the wall driven holographic quenches worth further studying.

\section*{Acknowledgements}
We want to thank J. Ignacio Cirac, Roberto Emparan, Karl Landsteiner, Fernando Marchesano, Erik Tonni and Tadashi Takayanagi for discussions. We very especially thank  Boye Buyens, Jutho Haegeman, Karel Van Acoleyen and Frank Verstraete for sharing with us some of their results prior to publication.
The work of E.daS. is financed by the spanish grant BES-2013-063972.
E.L. has been supported by the spanish grant FPA2012-32828 and SEV-2012-0249 of the Centro de Excelencia Severo Ochoa Programme. The work of J.M.  is supported in part by the spanish grant  FPA2011-22594,  by Xunta de Galicia (GRC2013-024), by the  Consolider-CPAN (CSD2007-00042), and by FEDER. A.S. is supported by the European Research Council grant HotLHC ERC-2011-StG-279579 and by Xunta de Galicia (Conselleria de Educaci\'on). Part of the numerical calculations were performed at the Centro de Supercomputaci\'on de Galicia (CESGA).


\end{document}